\documentclass{ab}

\usepackage{graphicx}
\usepackage{natbib}

\def\degr{\hbox{$^\circ$}}

\begin{document}

 \title{Polarimetric and Spectroscopic Observations of a Dynamically New Comet C/2012\,J1 (Catalina)}

\author{
{O.}~{Ivanova}$^1$, {O.}~{Shubina}$^2$,  A. ~{Moiseev}$^3$, V.~Afanasiev$^3$
}

\institute{
$^1$Main Astronomical Observatory, National Academy of Sciences, Kyiv, 03680 Ukraine\\
$^2$Taras Shevchenko National University of Kyiv, Kyiv, 01601 Ukraine \\
$^3$Special Astrophysical Observatory, Russian Academy of Sciences, Nizhnij Arkhyz, 369167, Russia}

\titlerunning{Polarimetric and spectroscopic observations of C/2012\,J1~(Catalina)}
\authorrunning{Ivanova et al.}

\date{January 5, 2015/May 27, 2015}
\offprints{Oleksandra Ivanova \email{sandra@mao.kiev.ua}}

\abstract{ We present the analysis of the polarimetric and
spectral data obtained for the dynamically new comet C/2012\,J1
(Catalina) when it was at a distance of 3.17~AU from the Sun. The
observations were made at the prime focus of the 6-m BTA telescope
using the SCORPIO-2 focal reducer. The map of the distribution of
linear polarization in the cometary coma was constructed. The
calculated value of linear polarization was on average about
$-2\%$. Spectral analysis of the cometary coma allowed us to
detect the emission of the CN molecule in the (0--0) band. The gas
production rate was derived using the Haser model and amounted to
$3.7\times10^{23}$~molecules per second. }

\maketitle

\section{INTRODUCTION}

The comets, minor planets, meteoroid particles, and interplanetary
dust form an evolutionarily interconnected family of small bodies
of the Solar System. Despite their modest masses, research of
these objects is very important from the cosmogonic point of view,
namely, to uncover the origin of the Solar System.  According to
modern cosmological ideas, comets are the relics of the Solar
System, and their composition should be most similar to that of
the primordial protoplanetary cloud. The question  about
parameters of dust particles (their size, orientation, structure)
in comas and tails of comets  active at large heliocentric
distances and the dynamically new comets, which  for the first
time enter the inner regions of the Solar System, remains
unsolved. At  heliocentric distances greater than 3~AU, the matter
is carried out from the surface of the cometary nucleus due to
sublimation of water ice and more volatile components (CO,
CO$_{2}$), the characteristic and outflow velocity of matter are
also varying. The question of variation of dust scattering
properties with heliocentric distance of comets also remains open.
Available observational data indicate a difference between
activity of dynamically new comets (active at large heliocentric
distances) and short-period
comets~\citep{Epifani2007:Ivanova_n,Epifani2008:Ivanova_n,Korsun2008:Ivanova_n,Korsun2010:Ivanova_n,Korsun2014:Ivanova_n}.
Hence, it is possible that the nature of  dust particles differs
as well.

Observations of comets using different methods (photometry,
spectroscopy, and polarimetry) and at different heliocentric
distances and phase angles provide useful information on the
material of the comet surface layer: the particle sizes, their
structure, and chemical composition. In this paper, we study a
dynamically new comet C/2012\,J1 (Catalina) using polarimetric and
spectroscopic methods.

Comet C/2012\,J1 (Catalina) was discovered on May 13, 2012 as  a
$16\fm4$ object within the Catalina Sky
Survey~\citep{Apitzsch:Ivanova_n}, a project searching for new
comets and asteroids. The comet revealed significant activity at a
distance larger than 3~AU. It  passed the perihelion on December
7, 2012 at a distance of 3.15~AU. The comet belongs to the class
of long-period comets and has a parabolic orbit with eccentricity
$e=1.001$  and
 orbit inclination   $i=34\fdg1$.

\begin{table*}[]
 \caption{\centerline{Log of Comet C/2012\,J1 (Catalina) observations}}
   \medskip
  \begin{tabular}{c|c|c|c|c|c|c|c}
  \hline
   Start time, & $r$,  & $\Delta$,  & $\alpha$,  & ${\rm PA}$,  & Grating/filter & Total exposure,  & Data \\
    UT &  au &  au &  deg  &  deg &  & s &  \\
  \hline
  Nov 15.8026, 2012 & 3.17 & 2.45 & 14.02 & 87.66 & $V$ & 15 & image \\
  Nov 15.8092, 2012 & 3.17 & 2.45 & 14.02 & 87.66 & VPHG1200@540 & 1500 & spectrum \\
  Nov 15.8280, 2012 & 3.17 & 2.45 & 14.02 & 87.66 & $V$ & 640 & image/polarization \\
  \hline
  \end{tabular}
\end{table*}

\section{OBSERVATIONS}

Our polarization and spectroscopic observations were carried out
on November 15, 2012, when the comet was at a distance of 3.17~AU
from the Sun and 2.45~AU from Earth, and its integral magnitude
was $13\fm9$.  The phase angle of the comet was~$14^{\circ}$,  its
scale in the sky plane was approximately  $1800$~km/arcsec. The
observations were performed on the  6-m BTA telescope of the
Special Astrophysical Observatory of the Russian Academy of
Sciences (SAO RAS) using the multimode SCORPIO-2 focal
reducer~\citep{Afan_Moi:Ivanova_n} operating in the spectroscopic
and polarimetric modes. An E2V\,42-90 CCD sized
$4600\!\times\!2048$ pixels was used as a   detector. The size of
the field of view was $6\farcm1\!\times\!6\farcm1$, the scale in
the $2\!\times\!2$ hardware  binning mode amounted to $0\farcs18$
per pixel.

Polarimetric observations were carried out in the broadband $V$
filter of the Johnson--Cousins photometric system. A polarizing
dichroic filter (Polaroid) was used as a polarization analyzer.

\begin{figure}
\includegraphics[width=\columnwidth]{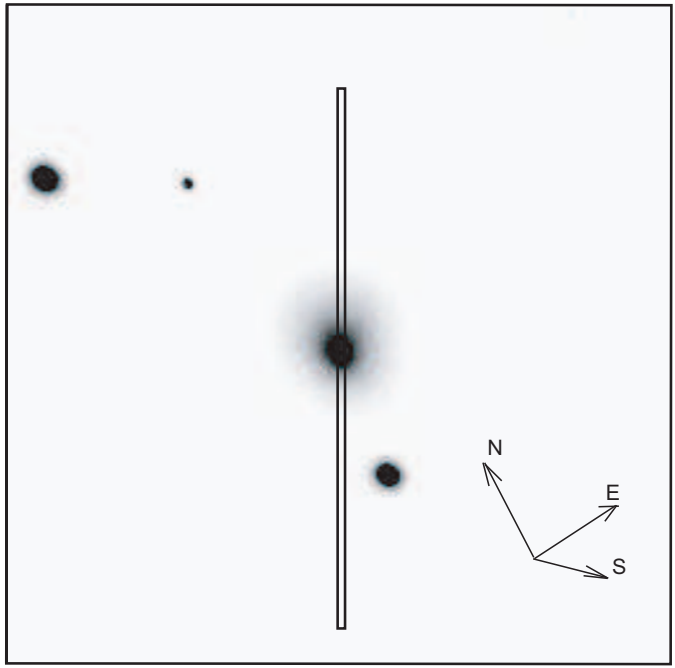}
\caption{An image of Comet C/2012\,J1
(Catalina) in the  $V$  filter  and the projection of the
spectrograph slit on the cometary coma  at the time of
observation. The directions to north, east, and to the Sun are
marked.} \label{fig1:Ivanova_n}
\end{figure}

Spectroscopic observations were carried out in the long slit mode.
The height of the slit amounted to  $6\farcm1$, the slit width was
1$''$. The grism with a volume-phase holographic grating
VPHG1200@540 provided the wavelength range of 3600--7200~\AA\ with
a spectral  resolution of about 5.2~\AA.  The observing log is
given in the table. Figure~\ref{fig1:Ivanova_n} shows a $V$-band
image of the comet and the projection of the spectrograph slit
onto the cometary coma.

\section{POLARIMETRIC INVESTIGATION}

To estimate the degree of linear polarization of the comet, we  used  the
 dichroic polarization analyzer installed in the spectrograph.
The analyzer was positioned at the three fixed angles:
$0^{\circ}$~and~$\pm60^{\circ}$. We obtained the intensity values
for these angles $I(x, y)_{0^{\circ}}$, $I(x, y)_{-60^{\circ}}$,
and $I(x, y)_{+60^{\circ}}$ and estimated the  Stokes $Q^{\prime}$
and $U^{\prime}$ parameters at each point of the image:
\begin{equation}
\label{form1:Ivanova_n}
  \left\{
   \begin{array}{ccl}
    Q^{\prime} &=& \frac{2I(x,y)_{0\degr} - I(x,y)_{-60\degr} - I(x,y)_{+60\degr}}{I(x,y)_{0\degr} + I(x,y)_{-60\degr} + I(x,y)_{+60\degr}}\,, \\[+15pt]
    U^{\prime} &=& \frac{\sqrt{3}}{2}\frac{I(x,y)_{+60\degr} - I(x,y)_{-60\degr}}{I(x,y)_{0\degr} + I(x,y)_{-60\degr} + I(x,y)_{+60\degr}}\,.
   \end{array}
  \right.
\end{equation}

The true values of the   Stokes $Q$ and $U$ parameters were
obtained from the following equations:
\begin{equation}\label{form2:Ivanova_n}
   \begin{array}{l}
    U = U^{\prime}\cos2\varphi - Q^{\prime}\sin2\varphi\,, \\[+5pt]
    Q = U^{\prime}\sin2\varphi + Q^{\prime}\cos2\varphi\,.
   \end{array}
    \end{equation}

To determine the degree of polarization $P$ and the angle of the
polarization plane  $\varphi$,  we used the following relations:
\begin{equation}\label{form3:Ivanova_n}
  \left\{
   \begin{array}{lcl}
    P &=& \sqrt{Q^{2} + U^{2}}\,, \\[+15pt]
    {\rm PA} &=& \frac{1}{2}\arctan\frac{U}{Q}\,.
   \end{array}
  \right.
   \end{equation}

Reduction of polarimetric images included subtraction of the bias
frame, flat field correction, and formation of images prepared for
processing. Next, using  the central isophotes of the comet image,
all the images were reduced to a single photometric center. To
improve the signal-to-noise ratio, the corrected   frames with the
comet image   were added up  using the median averaging procedure.
A detailed description of the technique of observation and data
reduction with the SCORPIO-2 spectrograph in the polarimetric mode
is described in~\citet{Afan_amir:Ivanova_n}.

\begin{figure}
\includegraphics[width=\columnwidth,bb=10 20 395 415,clip]{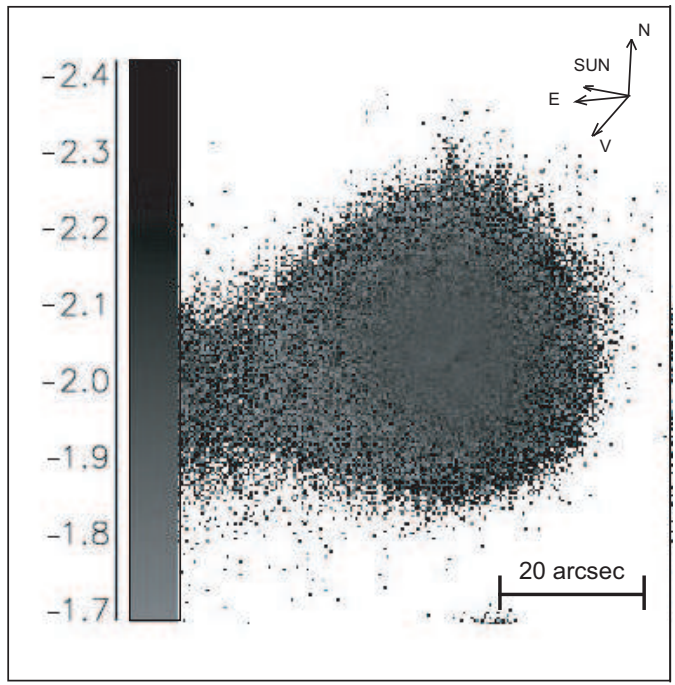}
\caption{Distribution of linear polarization
over the coma, projected onto the scattering plane. The scale of
the image is marked.} \label{fig2:Ivanova_n}
\end{figure}

Using formulas (\ref{form1:Ivanova_n})--(\ref{form3:Ivanova_n}),
we obtained
 the degree and angle of linear polarization for Comet
C/2012\,J1 (Catalina). Figure~\ref{fig2:Ivanova_n} presents the
distribution of linear polarization over the coma, projected onto
the scattering plane.

\begin{figure}
\includegraphics[width=\columnwidth]{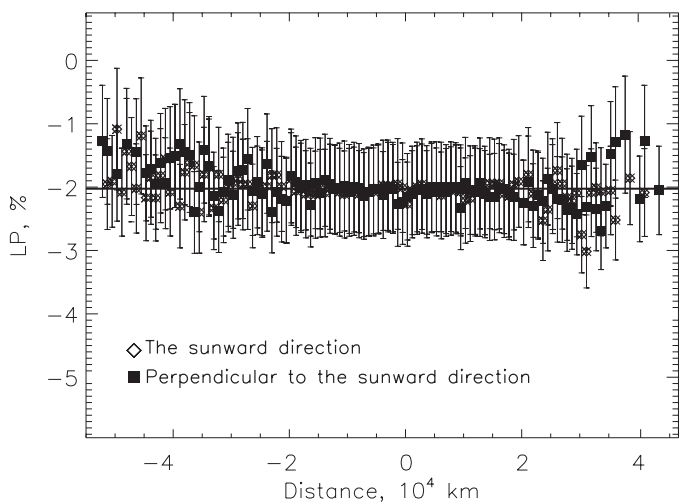}
\caption{Cross sections of the linear
polarization of the comet (LP).} \label{fig3:Ivanova_n}
\end{figure}

We have also made  sections  of the linear polarization of the
comet in the direction to the Sun and in the direction
perpendicular to it (Fig.~\ref{fig3:Ivanova_n}). The sections were
made through the photometric center of the image.

According to our estimates, the linear polarization over the
cometary coma (inside the   $\rho<17''$ radius, which corresponds
to approximately 30\,000~km) is on average $-2.03\%\pm0.1\%$. The
map of spatial distribution and the demonstrated cross-sections
(Figs.~\ref{fig2:Ivanova_n}~and~\ref{fig3:Ivanova_n}) show that
the nuclear region (up to 15\,000 km) reveals no abrupt changes in
the value of linear polarization. We do not consider the area
smaller than 5000~km, which is poorly resolved. In addition, there
may also occur errors due to the overlap of different images. In
the field with a radius of more than 25\,000~km, we possibly
observe  (in the absolute values) a small decrease of the linear
polarization.

\section{EXAMINATION OF THE COMETARY~SPECTRUM}

Primary reduction of  spectral data was performed using the codes
developed at SAO RAS, operating in the   IDL environment. It
included the bias frame subtraction, removal of cosmic ray traces,
flat field correction with the spectrum of a built-in lamp  with a
continuous spectrum,
 spectral line curvature compensation, wavelength calibration, and representation
of the data in an evenly spaced wavelength scale. The wavelength
calibration of the spectra
 was carried out using a \mbox{He-Ne-Ar} lamp.

To convert the cometary spectra into absolute power values, we
used the spectra of the standard star
BD\,+28$\degr$4211~\citep{Oke:Ivanova_n}, which was observed at the
same night. The response curves of the Earth's atmosphere for  SAO
RAS from the paper of Kartasheva and
Chunakova~\citep{kartasheva:Ivanova_n} were applied.

To analyze  the energy distribution in the cometary spectrum  and
search for possible molecular emissions, a one-dimensional
spectrum was built by adding up the counts along the slit within
$\pm18''$ from the nucleus.

\begin{figure}
\includegraphics[width=\columnwidth]{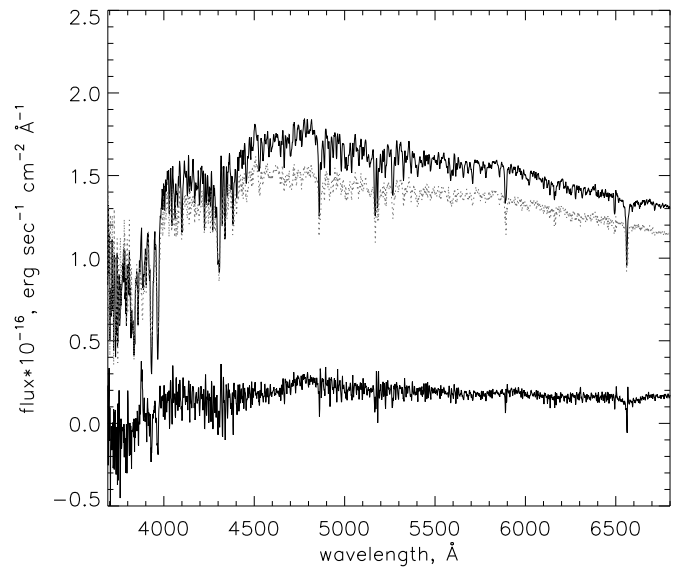}
 \caption{The spectrum of the comet (the
black curve), the modified solar spectrum (the bright dashed
line), and the result of its subtraction from the initial
spectrum.} \label{fig4:Ivanova_n}
\end{figure}

To isolate the emission spectrum of the comet, we subtracted from
the cometary spectrum  the high-resolution solar
spectrum~\citep{Neckel:Ivanova_n}, which had previously been
convoluted with a Gaussian function with an ${\rm FWHM}$
corresponding to the width of the instrumental profile in our
observations. The solar spectrum was scaled the way  that its
level was as close as possible to the low limit of the cometary
spectrum in the spectral intervals dominated by the continuum. In
addition, we introduced a correction  for the reddening effect.
The scale of the observed wavelengths was corrected for the
Doppler shift of the cometary spectrum (radial velocity of about
22~km/s). The qualitative analysis of the cometary spectrum
revealed the presence of weak molecular emissions. The total
spectrum of Comet C/2012\,J1 (Catalina) is presented in
Fig.~\ref{fig4:Ivanova_n}.

\begin{figure}
\includegraphics[width=\columnwidth]{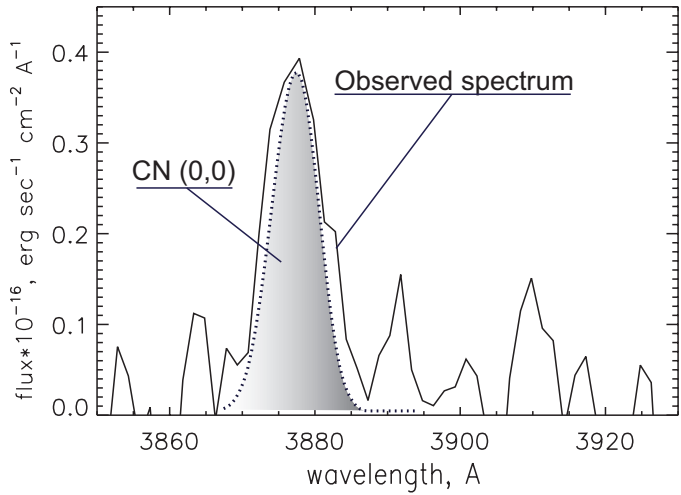}
\caption{Calculated and identified profile
of the CN molecule (0--0)  band.} \label{fig6:Ivanova_n}
\end{figure}

Identification of molecular emissions, observed in  comets in this
spectral range, was carried out by comparing the spectrum of the
comet with the calculated and laboratory spectra of molecules. The
spectrum reveals the presence of molecular emissions as well as
differences in the energy distribution of the cometary and the
solar spectra. Although the  blue spectral  region  is very noisy,
we managed to identify several spectral features which belong to
the CN  molecule  (electronic transition  \mbox{${\rm
B}^{2}\Sigma^{+}$--${\rm X}^{2}\Sigma^{+}$}).
 Since the rotational
 structure of the band in the resulting spectrum has not been resolved, we could
only identify  the edges  of the oscillating system (0--0) bands.
The theoretical spectrum for this molecule was built using the
LIFBASE code~\citep{Luque:Ivanova_n}, which is used to calculate
electronic spectra of diatomic molecules.
Figure~\ref{fig6:Ivanova_n} shows the identified spectral detail.

Assuming that the atmosphere of the comet is optically thin (in
this case, the fluxes are directly proportional to the number of
radiators), we can estimate the total number   $N(\rho)$ of
molecules contained in a column of  radius $\rho$ on the line of
sight using   the formula:
\begin{equation}\label{form5:Ivanova_n}
     N(\rho) = \frac{L}{g_{\lambda} }
     \,,
\end{equation}
where $L=4\pi\Delta^{2}F_{c}$ is the observed luminosity of the
comet in the emission band,  $F_{c}$ is the  flux of the comet in
this band,   $\Delta$ is the geocentric distance,  $g_{\lambda}$
is the efficiency of fluorescence for the given molecule at the
heliocentric distance
  $r=1$~AU,  which is determined by
the expression~\citep{ODell:Ivanova_n}:
\begin{equation}\label{form6:Ivanova_n}
     g_{\lambda} = \frac{\pi e^{2}}{m_{e}c^{2}}\,\lambda^{2}\left( \pi F_{{\rm Sun}\,\lambda} \right)f_{\lambda}\,w
     \,,
\end{equation}
where $m_{e}$ and $e$ are the mass and the charge of the electron,
$c$ is the speed of light, $f_{\lambda}$ is the oscillator
strength, $w$ is the probability of vibrational transition, $\pi
F_{{\rm Sun}\lambda}$ is   the density of solar radiation at
heliocentric distance $r$.

To convert the observed density of molecules in the  column
$N(\rho)$ into the molecule production  $Q$, we used the Haser
model~\citep{Haser:Ivanova_n}, according to which
\begin{equation}\label{form7:Ivanova_n}
\begin{array}{lr}
    \lefteqn{N(\rho) = \frac{Q}{v_{g}}\rho\left[ \int\limits_{x}^{x\mu}K_{0}(y)\,{\rm d}y\right.} & \\[+25pt]
   &  \left.  + \frac{1}{x}\left( 1 - \frac{1}{\mu} \right) + K_{1}(x\mu) - K_{1}(x) \rule{0pt}{24pt}\right]\,,
\end{array}
\end{equation}
where  $v_{g}$ is the speed of  molecule fly-out,
\mbox{$\mu=\gamma_{p}/\gamma_{d}$}, \mbox{$x=\gamma_{d}d$},
$\gamma_{p}$ and $\gamma_{d}$ are the characteristic path lengths
of the parent and the daughter molecules respectively, $K_{0}$ and
$K_{1}$ are the modified zero- and first-order Bessel functions of
the second kind.

Model calculations involve the full flux inside a specified
aperture. Since in our observations we only register the spectrum
from the regions captured by the slit of the spectrograph, a
corresponding correction needs to be done. When calculating the
aperture correction, we took into account both the ratio between
the slit areas and the circular aperture and the observed
brightness distribution along the slit.

For our calculations, we used the following values  of  model
parameters. The characteristic scales for the parent and daughter
molecules are  \mbox{$3.0\,r^{1.3}\times10^{4}$~km} and
$19.9\,r^{0.6}\times10^{4}$~km
respectively~\citep{Langland-Shula:Ivanova_n}.
 Since the \mbox{$g$-factor} for the
CN~(0--0) band is dependent on the heliocentric velocity of the comet,
 its value at the time of observation was refined using the
calculations made in~\citet{Schleicher:Ivanova_n}. In our case it
was \mbox{$2.4\times10^{-13}\,r^{-2}$~erg\,s$^{-1}$}  per
molecule. The rate of gas outflow was assumed to be the value
which is used in most of such calculations, 1~km
s$^{-1}$~\citep{Langland-Shula:Ivanova_n,Schulz:Ivanova_n}.

For the accepted values the CN gas production amounted to
$3.7\times10^{23}$~molecules/s.

\section{DISCUSSION}

Research of the properties of  gas and dust in comets with
different perihelion distances allows us not only to estimate the
possible differences or similarity of the cometary material
detected at different heliocentric distances but also to draw some
conclusions about physical conditions of comet formation in
general. Complex research of various comets, namely,  regular
monitoring of their spectra for the purpose of detection of gas
emissions, registration of  the onset of  formation of these
emissions as a function of heliocentric distance as well as
investigation of polarization distribution of   gas and dust in
the comets, gives the possibility of building a realistic model of
the cometary nucleus. The short-period comets and a number of
long-period comets with perihelia shorter than 2~AU, for which
 enough observed spectral, photometric, and polarimetric material was accumulated, are by now  the most thoroughly studied.
New observational strategies and the use of large telescopes has
recently allowed the active research of objects from distant
outskirts of the Solar System. Focusing on the research of
properties of dynamically new comets that arrive in the inner
Solar System  for the first time, we would make a comparative
analysis aimed to classify the groups of comets by physical
properties of dust and gas, which are directly linked to the
evolution and/or places of origin.

The aim of our polarimetric and spectroscopic observations was to
study a dynamically new comet C/2012\,J1 (Catalina). During the
analysis of polarimetric data, we measured the magnitude of linear
polarization.  Its average value over  the cometary coma varies
from $-1.9\%$ to $-2.1\%$. Polarization of  scattered radiation
depends on the heliocentric distance, phase angle, and morphology
of the coma, which is directly connected with the activity of the
comet. Many comets, including the distant ones, reveal jet
activity, the consequence of which may also be the variation of
polarization properties of the dust particles of the coma. Due to
the jets, the nuclear region
 may reveal areas with different polarization.

Polarization is also dependent on the diameter of the aperture
within which the measurements were made. The aperture dependence
appears in general case due to different contributions of the
continuum  and the gas emissions into the integral radiation. For
this reason, we can only compare polarization in different comets
(assuming equal remaining parameters: similar fields, cut by the
filters, comets without emissions) in cases of similar sizes of
the measured coma. In our study, we did not use the apertures for
observations, and  the broadband filter that we applied did not
capture any bright molecular emissions.

The constructed polarization map allowed us to explore different
regions  of the cometary coma. The analysis of the distribution
map, complemented by the plotting of polarization sections
performed in the direction to the Sun and in the direction
perpendicular thereto, indicated that linear polarization does not
essentially vary over the cometary coma. Our observations do not
reveal any regions with different polarization. Based on this, it
can be assumed that the coma of Comet C/2012\,J1 (Catalina) has no
isolated jet structures as in Comet
29P/Schwassmann--Wachmann\,1~\citep{Ivanova:Ivanova_n}, or a
fan-shaped tail as in Comet 2P/Encke~\citep{Kiselev:Ivanova_n},
which can lead to the occurrence of regions with different
polarization.

If we compare the values of  linear polarization given
in the literature for the   dust comets,    our
 value proves to be the highest by the absolute value. For example, the degree of polarization was
 $-1.49\%$ (at a phase angle of  11$^{\circ}$) for Comet 1P/Halley,
$-1.09\%$ ($\varphi=14\fdg8$) for Comet  C/1989\,X1 (Austin),
$-1.4\%$ ($\varphi=12\fdg56$) for 47P/Ashbrook--Jackson, $-1.78\%$
($\varphi=15\fdg7$) for C/1990\,K1
(Levy)~\citep{rosenbush:Ivanova_n}.  But this difference in the
degree of polarization can be related to the fact that our value
was calculated for the entire coma, while those given for
comparison are for the apertures. However, in some distant comets
with perihelia of more than 4~AU which are active over long
distances and virtually without
emissions~\citep{Korsun2008:Ivanova_n, Korsun2010:Ivanova_n}, the
linear polarization was also higher than for the comets that were
observed at close distances from the Sun at the same phase
angles~\citep{Ivanova:Ivanova_n}. In this case, polarization
variations can be caused by the conditions that lead to activity
of the comet  at different distances from the Sun, as well as by
the rate of dust loss, which is also different for distant comets
and comets observed at a distance of less than 2~AU.

The spectroscopic observations of Comet\linebreak C/2012\,J1
(Catalina) registered the CN molecule emission   in the  (0--0)
band of its violet system. At a distance of more than 3~AU, most
of  the cometary spectra are poor in emissions. But in a number of
distant comets,
 emission features of the CN molecule were observed at large
heliocentric distances~\citep{Korsun2006:Ivanova_n,
Korsun2008:Ivanova_n, Korsun2014:Ivanova_n}. The cometary spectra
(e.g., for Comet C/1995\,O1 (Hale--Bopp)) at such distances also
reveal the C$_{2}$ molecule
emissions~\citep{a'hearn:Ivanova_n,Fink:Ivanova_n}, but they were
not identified in the spectrum of our comet.

Measuring the flux in the CN emission, we estimated the rate at
which these molecules arrive into the cometary coma. According to
our calculations, gas production amounts to
\mbox{$3.7\times10^{23}$}~mole\-cules/s for the CN molecule.
Comparing our results with the data
from~\citet{Langland-Shula:Ivanova_n},
 we can conclude that the CN molecule gas production is close to the lower limit of this
value for the majority of comets observed at heliocentric
distances closer than 2~AU., indicating  low
 gas production of the comet.

\section{CONCLUSION}

The analysis of polarimetric and spectral observations of the
dynamically new comet  C/2012\,J1 (Catalina), performed on the
SAO~RAS 6-m telescope, suggests the following conclusions:
\begin{list}{}{
\setlength\leftmargin{2mm} \setlength\topsep{2mm}
\setlength\parsep{0mm} \setlength\itemsep{2mm} }
 \item (1)  the average value of  linear polarization (projected onto the scattering plane) ranges from  $-1.9\%$ to $-2.1\%$;
 \item (2)  the spectrum of the comet revealed the CN molecule emission in the  (0--0) band of its violet system;
 \item (3)  gas production for the CN molecule is\linebreak \mbox{$3.7\times10^{23}$}~molecules/s;
\end{list}

\begin{acknowledgements}
The observations on  the 6-m  BTA telescope  were conducted with
financial support of the Ministry of Education and Science of the
Russian Federation (agreement no.~14.619.21.0004,  project ID
RFMEFI61914X0004). The authors thank P.~P.~Kor\-sun for critical
comments and discussion, as well as the anonymous referee for
comments that helped to improve the article. A.~Moiseev thanks the
non-profit {\it Dynasty} foundation for financial support.
\end{acknowledgements}

\end{document}